\documentclass[runningheads]{llncs}
\usepackage[T1]{fontenc}
\usepackage[T2A]{fontenc}
\usepackage[utf8]{inputenc}
\usepackage{hyperref}
\usepackage{graphicx}

\begin{document}
\title{``Flipped'' University: \\ LLM-Assisted Lifelong Learning Environment}
\titlerunning{LLM-Assisted Lifelong Learning Environment}


\author{Kirill Krinkin\inst{1}\orcidID{0000-0001-5949-7830} \and
Tatiana Berlenko\inst{1}\orcidID{0009-0001-8141-2095}}
\authorrunning{K. Krinkin, T. Berlenko}

\institute{Constructor University, Bremen, Germany \email{kirill@krinkin.com}\\
\and
JetBrains Ltd, Cyprus \email{tatiana.berlenko@gmail.com}
}
\maketitle              

\begin{abstract}
The rapid development of artificial intelligence technologies, particularly Large Language Models (LLMs), has revolutionized the landscape of lifelong learning. This paper introduces a conceptual framework for a self-constructed lifelong learning environment supported by LLMs. It highlights the inadequacies of traditional education systems in keeping pace with the rapid deactualization of knowledge and skills. The proposed framework emphasizes the transformation from institutionalized education to personalized, self-driven learning. It leverages the natural language capabilities of LLMs to provide dynamic and adaptive learning experiences, facilitating the creation of personal intellectual agents that assist in knowledge acquisition. The framework integrates principles of lifelong learning, including the necessity of building personal world models, the dual modes of learning (training and exploration), and the creation of reusable learning artifacts. Additionally, it underscores the importance of curiosity-driven learning and reflective practices in maintaining an effective learning trajectory. The paper envisions the evolution of educational institutions into ``flipped'' universities, focusing on supporting global knowledge consistency rather than merely structuring and transmitting knowledge.

\end{abstract}

\section{Introduction}

The last few years have witnessed an explosive development of technologies related to artificial intelligence and its applications. New algorithms, approaches, models, and services based on them appear every day. Even for domain experts, keeping track of these changes is almost impossible. The same happens in other domains where AI-assisted tools play crucial roles in research and development. 
Gaps in the knowledge and skills of individuals and professional groups are getting wider, and specialization is increasing. All this is happening against the rapid deactualization of applied knowledge and skills. Today, it is no longer enough to receive a fundamental general education at university and then, in career life, only to improve professional mastery. Many studies support the inevitability of a shift to continuous lifelong learning\cite{Blossfeld2019,bjet.13121,bjet13123}. However, much of this research is positioned within the framework of \emph{androgogy}~\footnote{The method and practice of teaching adult learners; adult education}, as a complement to the core education received at school and university. 

Despite the historical traditions of building educational systems that separate school, university, and post-university education levels, a unified model should be considered. The main reason for the historical separation of educational systems is the institutionalization of the learning environment: they have their knowledge corpus to teach and a methodology. In addition to teaching methodology, one of the primary responsibilities of educational institutions is to ensure coherence and structuring of knowledge within the curriculum. Nevertheless, 
nowadays, any knowledge that appears in the public space becomes instantly available to anyone, and tools based on Large Language Models~(LLM) can automatically provide an initial structure for the knowledge (with not ideal quality, but good enough for preliminary exploration). Thus, universities are losing the function of structuring knowledge. Moreover, the interdisciplinary nature of knowledge plays against any ``correct structure'' because this structure depends on concrete context (and, quite often, the particular problem). 

From the learning experience perspective, active structuring itself (guided by a tutor or independent)  significantly increases the speed and quality of learning material acquisition. In this case, the learner is primarily responsible for assimilating new knowledge~\cite{Bloom1968,Morris2020,Kolb2015,Hou2017}, and the learning environment should enforce such an activity. This reveals a transformation trend from institutionalized educational institutions to a personal, self-driven, and self-designed, lifelong education system. Ideally, everybody could have a personal self-constructed ``university'' that is well-tuned for career trajectory and natural inclinations. 

This paper describes a conceptual framework (principles and components) for building a self-constructed lifelong learning environment. Many aspects of education (psychological, cultural, methodological, age, and other) are not the subject of this paper. Also, the modern technologies of multimodal interaction between a human and a computer, based on the generation of sound, video, and other signals (which, of course, is a promising area of research) are not discussed. The focus is mainly on the natural language interface in the text form.

It is important to note that we rely on the idea that education is based on three fundamental processes: individual study, collaborative study, and sharing learning experiences. The first process involves self-directed learning, wherein individuals independently acquire new knowledge and skills. The second process is joint co-learning, which encompasses (often in project-based form) activities where people work in one consolidated learning environment within a shared space. Here, groups of learners collaboratively address tasks, thereby acquiring new knowledge and sharing experiences during the course of their joint work. The third process involves sharing one's learning experiences with others. This entails disseminating artifacts created during the individual study. This paper focuses mainly on the first process (individual study) without diminishing the importance of other components.

\section{State of the Art}

It is necessary to consider the broader context to discuss the application of LLMs to construct personal learning environments. This section provides an overview of the role of natural language in learning and the genesis of computer-aided learning tools.

\subsection{Language in Learning}

Language (whether written or spoken) is at the heart of human learning and understanding. It is a tool for accessing existing knowledge and forming new concepts and meanings. Learning is in humans' nature. They evolved to specialize in the cognitive niche, which is deﬁned by reasoning about the causal structure of the world, cooperating with other individuals, sharing that knowledge, and negotiating those agreements via language~\cite{pinker2010}.
B.~Bloom claimed that the student's ability to understand instruction is primarily determined by verbal ability and reading comprehension~\cite{Bloom1968}. This means that mastery of the language is essential for success in learning. Many research supports~\cite{Christiansen2008,Fausey2010,Schoenemann} that the evolution of the human brain, knowledge, and language are incredibly highly correlated. They are all part of the joint co-evolutionary process. 

Before the emergence of large language models, many ways of representing knowledge and related information (e.g., relational, object databases, ontologies) and their corresponding access interfaces, including languages (e.g., Prolog, SQL, OWL, etc.), were created. However, having an artificial language or tool to access knowledge has the following limitations.
\begin{itemize}
    \item High \emph{learning curve}~\footnote{Hermann Ebbinghaus introduced this term to measure how much effort is required to understand (memorize) new knowledge.}. Suppose access to knowledge requires mastery of an artificially created tool or language. In that case, initial knowledge is needed to learn the tool, and the learner may not have enough of it.
    \item Context Specificity. The expressive power of any artificial language strongly depends on the context of the application. For example, the language of mathematical formulas is well suited for describing physical processes and computational problems but will hardly be convenient for describing historical events, philosophical concepts, and art.
\end{itemize}

With the advent of LLMs, it is technically possible to interact with the entire body of accumulated knowledge in natural language. This practically removes the high entrance barrier and context dependency. In learning, the learner's natural language evolves to acquire new concepts. The learner becomes capable of accessing more specific and complex knowledge. LLMs can be considered as a navigation tool through comprehensive knowledge graphs. It was initially generated by humans and consolidated by machines. Probabilistic edges connect concepts based on human \emph{consensus} implicitly expressed in the native and artificial languages~\cite{krinkin2024}. In other words, humans transfer knowledge through natural language, and artificially created languages help to make this transfer more efficient in particular domains or contexts.

\subsection{Computers in education}

The idea of using computers and software to teach people has existed since almost the day computers appeared. This section describes the genesis of approaches to using language and texts.

\subsubsection{Teaching machines and programmed learning.} 

In the 1960s, the concepts of \emph{teaching machine} and \emph{programmed learning} were actively discussed. The main idea was as follows~\cite{lumsdaine1960}. The learning material (or, literally, textbook) had to be broken down into small blocks -- \emph{frames}. These are combined (programmed) into programmed learning sequences. The computer should progress from the learner's initial knowledge\footnote{In the original paper, authors used ``behavior'', substitution it by ``knowledge'' or ``skill'' does not change the general meaning } in small steps through the development of more complex knowledge and skills. This progression can occur in small enough steps so frequent failures do not jeopardize the student's progress and motivation. A central process for the acquisition of knowledge is reinforcement. Knowledge is acquired under conditions in which a response is followed by a subsequent ``rewarding condition''. 

Later, it became clear that, on average, any artificial framing (splitting of learning material into fragments) and then forming (programming) learning chains do not give a noticeable gain in learning. The reason is that when learning the material, the learner does not move linearly through the text. He uses not only the text but also the structure. S.~Pressey claims that conventional textbooks order and structure their contents in paragraphs, sections, and chapters for a learner with reading-study skills. They exhibit that structure in headings and the table of contents, making all readily available in an index with page headings and numbers. The learner thus has multiple aids to develop and structure \emph{his own} understanding. He can skip the already known, turn back due to a later felt need, and review selectively. The best [knowledge structure representation] is closer to texts than programs\footnote{The text in brackets is ours}~\cite{pressey1963}. Each learner will read a textbook with his unique pattern, switching attention between frames according to current understanding and prior knowledge.

\subsubsection{Massive open online courses.} Until the 2000s, there were limitations in the university educational system, such as high tuition costs and geographical barriers. To avoid this problem, the most prominent universities (MIT, Harvard, Stanford, etc.) began to publish educational materials into the open access\footnote{The first ``OpenCourseWare'' (OCW) initiative was founded by MIT in 2001.} and an idea of social learning (students act as a community and help each other through discussion forums) appeared~\cite{brown_adler_2008}. In 2008, the first Massive Open Online Courses (MOOC) were published, providing free educational opportunities for anyone with an Internet connection. Technically, the first MOOCs were theoretical materials, such as recorded video lectures or text, and automatically verifiable tasks (tests). MOOCs have been quite successful in solving the problem of accessibility of learning. However, the idea of social learning, which envisioned students helping each other and learning together, has failed. Despite the availability of discussion forums, with an average enrollment of 43'000 students, only 6.5\% of students complete the course~\cite{Jordan2014}. 

\subsubsection{Massive Adaptive Interactive Text.} 
The main problem with MOOCs is that they focus on the masses while sacrificing the individuality of the approach to the particular learner. MOOCs have focused on mass appeal while they should strive for individuality. Nevertheless, misunderstanding is deeply individual. 
%
The student's limited ability to ask questions causes gaps in understanding to accumulate. When misunderstandings reach a critical mass, they stop the possibility of learning. 

In 2015, P.~Pevzner and P.~Compeau proposed to use the ``flipped'' classroom approach~\cite{bergmann2012flip} for MOOCs, where students must watch short video modules and work through the Massive Adaptive Interactive Textbook (MAIT)~\cite{Pevzner2015}. Such an approach has been implemented for a course on bioinformatics algorithms and tested over several years. The idea is the same as \emph{programmed learning}, but with the following significant modifications:
\begin{itemize}
    \item instead of a fixed pre-programmed study trajectory, students have multiple possible paths between topics (steps); the learning environment is adapting to student performance and understanding by changing this trajectory on the fly;
    \item traditional quizzes are replaced by programming mini-challenges. Authors acknowledge the limitation of that type of challenges for non-STEM\footnote{STEM stands for Science, Technology, Engineering, Mathemathics} domains;
    \item authors constantly update the course by adding new pieces of the learning material with alternative explanations for the same concepts based on students \emph{learning breakdowns}. It converts traditional textbooks to <<hyper-textbook>> -- the book that can be read in different ways.
\end{itemize}

Using the ``flipped'' classroom approach implies that the student learns the material independently and tests his knowledge with automatic tools. When learning breakdowns occur, a topic that could not be mastered independently, the student can discuss the problem with the teacher (ideally individually or in a small group). The capstone projects can help students practice applying new knowledge in a broader context. 

\subsubsection{LLM-Assisted Learning} 

The success of Large Language Models~\cite{minaeeLargeLanguageModels2024} (such as the GPT, LLaMA, PaLM, etc.) in generating texts has created an explosion of interest in their application in education. They are applicable at all levels of education: elementary school, middle and high school, university, and professional education. They have great potential for different formats: individual and in group, on-line and off-line. They are suitable for learners and tutors to create content, to support the learning process, for knowledge assessment and evaluation etc~\cite{AYANWALE2022,KASNECI2023102274}.

The application of LLMs to education is relatively new, and today, there is no consensus or best practice yet on a wide range of issues, such as models of learning, security and privacy, copyrights, ethics, credibility, abilities to explain, training bias, and many more. However, the main potential of LLMs for education lies in their crucial ability to adapt the learning environment to the individual. By leveraging the power of data-driven personalization, timely feedback, and dynamic content delivery, adaptive learning systems can enhance student engagement, foster self-directed learning, and improve overall learning outcomes~\cite{Gligorea2023}.

\section{Lifelong Learning Principles}

Humans live in a complex, dynamic environment -- \emph{complex system}~\footnote{\href{https://en.wikipedia.org/wiki/Complex_system}{https://en.wikipedia.org/wiki/Complex\_system}}. To act effectively, they need to have knowledge about this environment to adapt. Building a deterministic (or computable) model is fundamentally impossible due to the nature of such systems. They are composed of many components which may interact with each other. 
In the process of evolution, humans invented language, a \emph{complex \textbf{adaptive} system}~\cite{Christiansen2009,Schoenemann}, for knowledge accumulation and sharing. It can be argued that \emph{learning} and \emph{adaptation} are parts of the same process of evolution. 

\begin{figure}
\includegraphics[width=\textwidth]{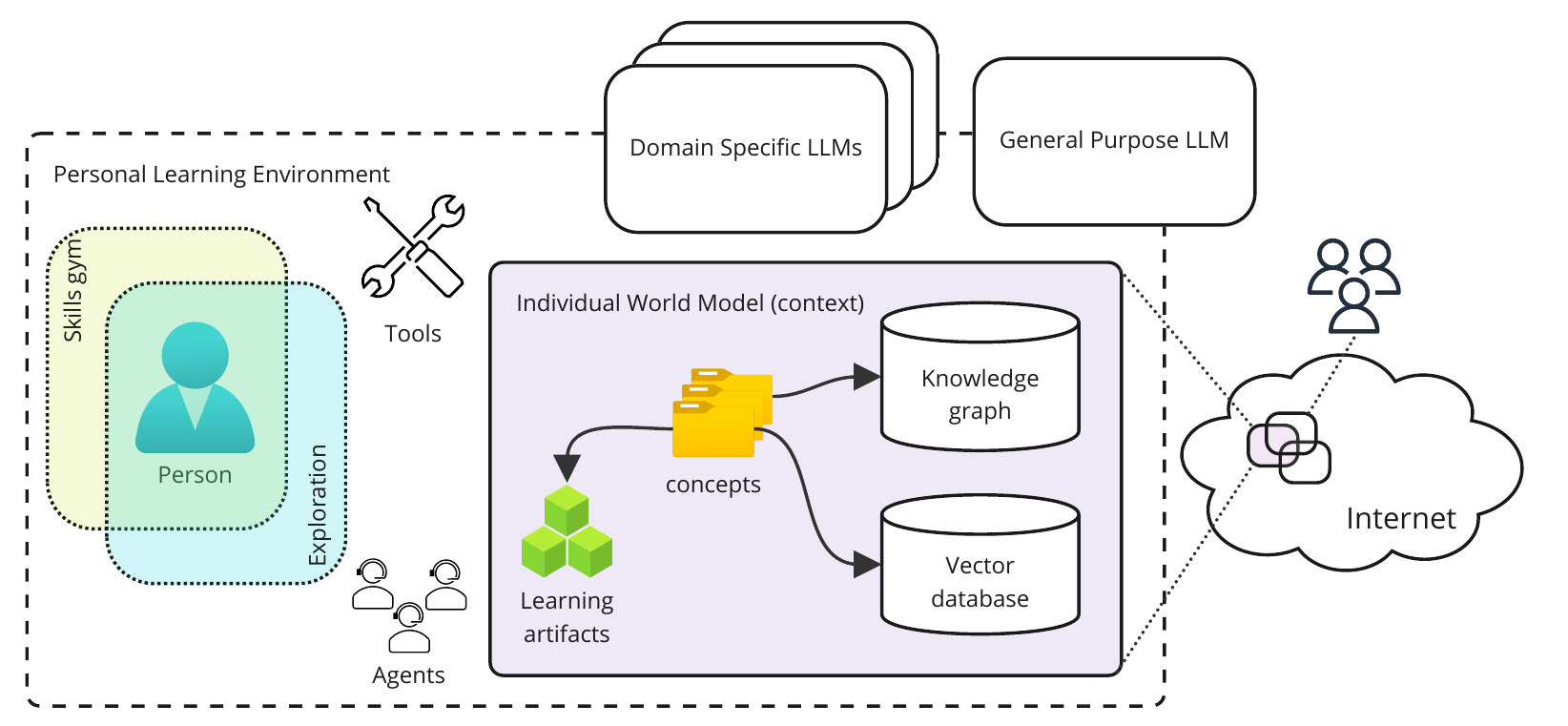}
\caption{LLM-Assisted knowledge acquisition.}\label{ple}
\end{figure}

Being a model of language, LLMs are consequently models of the world and the prime mediators of knowledge (see Fig.~\ref{ple}). A learner can navigate through all existing knowledge by using general purpose (like the ChatGPT) and domain-specific LLMs.

The speed of change in the environment is increasing, requiring an ever greater speed of adaptation and, thus, more effective lifelong learning. The following are lifelong learning principles.

\begin{itemize}
    \item Learners should build their \emph{own, personal, world model} (body of knowledge). 
    
    \item The Learning process has two modes -- \emph{training} (skill gym) and \emph{exploration}. Those modes correspond to System 1 (fast, automatic, and intuitive) and System 2 (slow, conscious, and symbolic), introduced by D.~Kahneman~\cite{kahneman}. In the exploration mode, learners explore new concepts and adapt them to their needs (general purpose and domain-specific LLMs mediate between human and global bodies of knowledge). In the training mode, learners master routine tasks with training software to build new unconscious behavior patterns related to new knowledge.
    
    \item The learning process creates \emph{reusable learning artifacts}. This synthesis is essential for deeply understanding concepts and future knowledge evolution~\cite{bloom2001,Engelbart1962}.

    \item In active lifelong learning, the \emph{personal intellectual agency} is created and trained. It consists of LLM-based agents who assist in various aspects of knowledge acquisition. The agents are active participants in the learning process. Their skills evolve with the individual world model. Artifacts of interaction with intelligent agents (e.g., dialogues, instructions, the result of data processing) become part of this model.

    \item Learning should be \emph{curiosity-driven}. Each learning session starts with a practical problem or open question and a few initial related concepts to study. Those concepts can be suggested by a tutor, an intellectual agent, or chosen by the learner according to their problem interpretation. There should be no direct instructions on what and how should be learned. Switching between training and exploration processes, the learner eventually will be deep enough to create comprehensive learning artifacts. The personal intellectual agency could contain agents that will help to maintain the required level of curiosity~\cite{Abdelghani_2023}.

    \item \emph{Reflection} is a fundamental process guiding the direction of lifelong learning.
    With their learning environment (including the body of knowledge, personal intellectual agency, and learning artifacts), the learner is a united complex system -- evolving hybrid\footnote{A system consists of human and artificial agents} intelligent organism. 
    Learners periodically analyze their performance, recognize good and bad learning patterns, and upgrade agents' instructions and data accordingly. Personal learning environment and experience itself is the subject of lifelong study. 

    \item \emph{Shared learning experience} is essential for enriching a personal world model. It should be based on solving open problems (or comprehensive games) with other learners. It should be a challenge for learners to actively apply and improve knowledge and skills acquired in personal study. It can be done with \emph{chaordic learning}~\cite{krinkin2017} approach, which allows to facilitate such activities in offline and online modes~\cite{krinkin2022}.

\end{itemize}


The main advantage of personal educational environments built according to listed principles is a holistic approach and total personalization. Learners become owners, creators, and maintainers of their own knowledge acquisition system. Using the ``flipped'' classroom metaphor~\cite{Pevzner2015}, the education system could be transformed into a personal lifelong university (``flipped'' university). In this scenario, the university network will not be so crucial for structuring and translating knowledge but will become indispensable in supporting the consistency of the global knowledge evolution process.



\section{Personal Lifelong Learning Environment}

\begin{figure}
    \includegraphics[width=\textwidth]{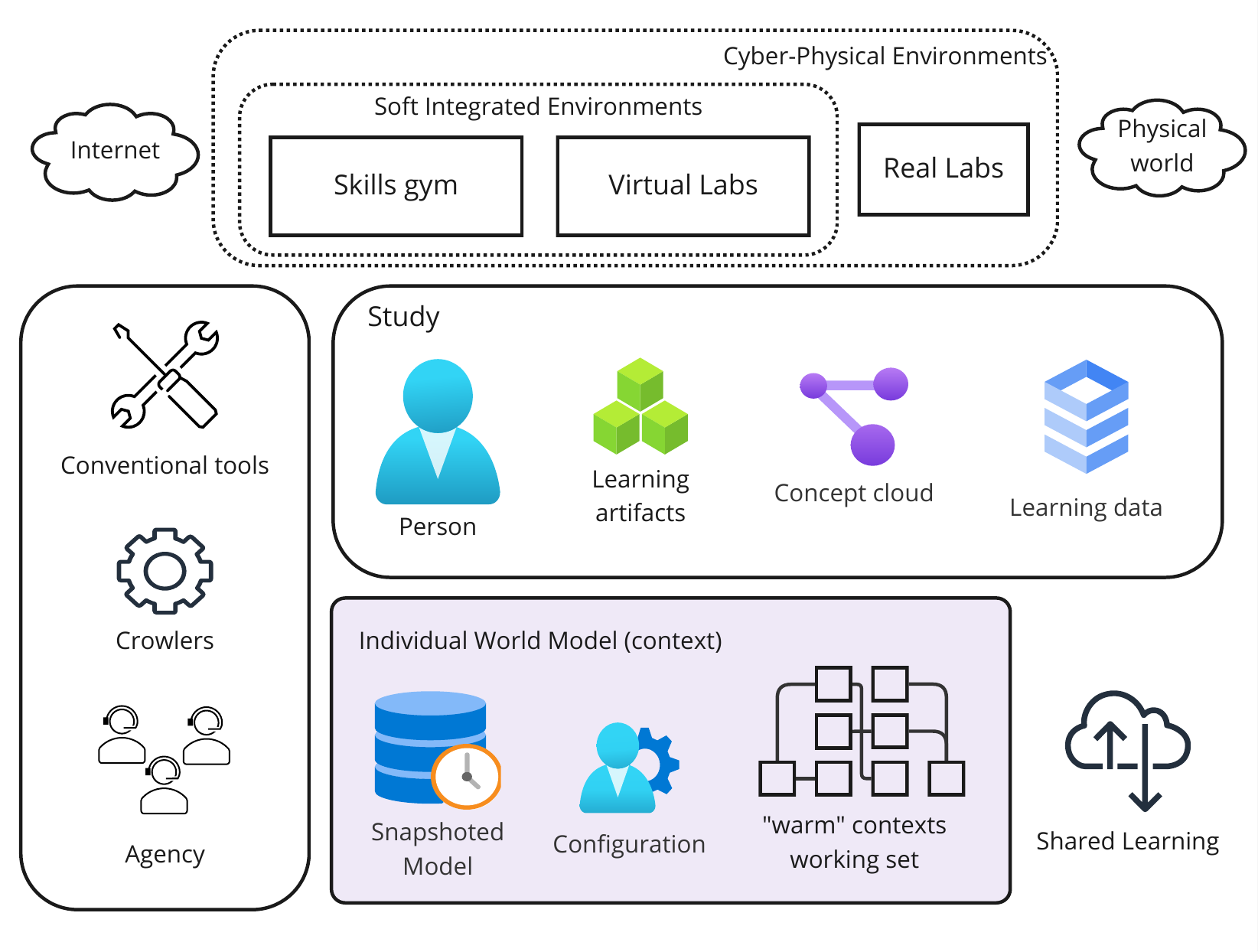}
    \caption{Personal Lifelong Learning Environment} \label{components}
\end{figure}

This section discusses the general components of a personal learning environment. It can be used as a reference model for future development (see Fig.~\ref{components}).  
The initial idea of creating the computer-aided environment for extending human abilities in study and research belongs to D.~Engelbart. He introduced a conceptual framework (operating system, OS) for augmenting human intellect~\cite{Engelbart1962}. The principal elements are the language artifacts and methodology a human has learned. These elements are dynamically interdependent within an operating system. The core feature of this system is the automation of the symbol manipulation associated with the minute-by-minute mental processes. It helps to organize the artifacts' basic human capabilities and functional capabilities, which are organized successively into ever more sophisticated capabilities.

Currently, Engelbart's concept is closest to the LLM OS architecture, introduced by A.~Karpathy~\cite{intro_to_llms_2024}, which has the following components (metaphorically referring to the conventional operating system). 
\begin{itemize}
    \item \emph{main process} (operating system core) -- Large Language Model;
    \item \emph{memory and knowledge} -- hard disk (persistent storage) containing dialog history, structured and unstructured data in files, and text embeddings 
    \item \emph{system utilities} -- classical applications like calculator, Python interpreter, text editor, terminal client;
    \item peripheral devices can input/output signals as video, audio, and text.
    \item browser as a connector to the Internet;
    \item connectors to other LLMs.
\end{itemize}

Using LLM OS design is a convenient way for \emph{personal lifelong learning environment} implementation. 

\subsection{Study objects}

The learning process focuses on working with the following objects: learning artifacts, concept cloud, and learning data.

\emph{Learning artifacts.} During the study of new concepts, reusable learning artifacts should appear. They are objects that allow learners to return to them after some time to recall the previous learning experience quickly. These can be texts (notes that arise while learning concepts), drawings with visualizations, results of computations or experiments, dialogues with intelligent agents, results of search queries, and developed programs. They can also be physical objects if the educational experience involves experiments in the real world.

\emph{Concept cloud.} Transformer-based LLMs are suitable for the initial structuring of concepts and navigation through a global knowledge system. They are not robust to text graph topology. In a conventional textual knowledge base, documents have semantic links. These links are as crucial for navigation aside from navigation based on text similarity. A concept cloud is a textual graph (hypertext) reflecting the links between concepts in the learner's knowledge system. It can be maintained manually (similar to Wikipedia) or generated in collaboration with LLMs in the Graph Retrieval-Augmented Generation (GRAG)~\cite{edge2024,hu2024} process. A node in such a graph is \emph{concept description}, representing only a name (a word) in the trivial case. In the learning process, learning artifacts, concept descriptions, and other objects from the learning environment and the Internet are associated with this node. Eventually, the personal body of knowledge is individual interpretations of concepts. In most cases, those interpretations are shared with a referent group.

\emph{Learning data.} Any data that is used for the study: factual, simulated, generated, or retrieved from internal or external databases. A particular case is the data generated by LLM-based intelligent agents from the personal body of knowledge, including trends related to the evolution of the individual learning environment (history of lifelong learning). 

\subsection{Tools and assistants}

The following tools support the learning process: conventional programs and services running on a local computer or the Internet (all good old software we use daily), crawlers, assistants, automated skills trainers, and simulators. 

\emph{Conventional tools.} Many tools and services already exist for structured data and information processing. To integrate them with LLMs, binding frameworks\footnote{like LangChain} can be used. They allow the combination of natural language instructions with conventional tools and connect all of them in sequences. For example, the learner can create in his learning environment a chain: extract information from an article in PDF format $\rightarrow$ search for available datasets in SQL format by keywords $\rightarrow$ execute an SQL query $\rightarrow$ generate a text report in a given format. The artifact created in this way can be a subject for study, correction, and verification.

\emph{Crawlers.} During the study, a learner requires different kinds of information about the current (being studied) concept cloud subset. Crawlers can automatically identify what should be provided and perform the search. For instance, the following type of search can be useful:
\begin{itemize}
    \item in previous learning experiences (or publicly available other learning artifacts on the same topics);
    \item existing learning systems and courses related to the topic;
    \item prime sources for the learning concepts;
    \item classic textbooks;
    \item last scientific results;
    \item items in public libraries and databases;
\end{itemize}

Learners should configure and tune those crawlers depending on the topic of study.

\emph{Agents.} The learner configures an agency of LLM-based agents for a specific study. They play as learning companions, creating the illusion of a ``collective educational experience''. For example, suppose the subject is a programming language. In that case, the following roles may be useful: trainer (generates and checks assignments), demonstrator (generates example programs and explains how they work), explainer (explains new material using analogies from domains known to the learner), ``younger brother'' (a less experienced colleague who asks the learner to explain elements of the learned material in simple language), critic (criticizes current solutions, points out opportunities for improvement). Such kind of collaboration between human agent and artificial agents corresponds to co-evolutionary hybrid intelligence paradigm~\cite{krinkin2021,krinkin2023}.

\section{Conclusion}

This paper has explored the transformative potential of LLMs in creating a personalized, lifelong learning environment. Fundamental principles identified include the necessity for learners to build and continuously update their own world models, the dual modes of training and exploration in the learning process, and the creation of reusable learning artifacts. Integrating LLM-based agents can also provide personalized guidance, fostering an active and engaging learning experience. 

It is important to note that the presented conceptual framework and principles are insufficient to define the future of education integrated with artificial intelligence. A massive practical evaluation is highly demanded. The learning models and the study success criteria may undergo significant changes. From this perspective, the LLM-Assisted Lifelong Learning Environment and provided principles can be considered an exploration and experimentation apparatus to shape future education.

The \emph{``flipped'' university} metaphor can demonstrate the shift from institutionalized education to a more flexible and adaptive system in which learners take ownership of their educational journeys. 

\section{Acknowledgement}

This preprint has not undergone peer review or any post-submission improvements or corrections. The Version
of Record of this contribution is published in \textbf{volume} and is available online at https://doi.org/~\footnote{References will be provided after publishing ICONIP-2024 proceedings}.

\bibliographystyle{splncs04}
\bibliography{7455}

\begin{thebibliography}{10}
\providecommand{\url}[1]{\texttt{#1}}
\providecommand{\urlprefix}{URL }
\providecommand{\doi}[1]{https://doi.org/#1}

\bibitem{Abdelghani_2023}
Abdelghani, R., Wang, Y.H., Yuan, X., Wang, T., Lucas, P., Sauzéon, H., Oudeyer, P.Y.: {GPT}-3-driven pedagogical agents to train children’s curious question-asking skills. International Journal of Artificial Intelligence in Education  \textbf{34}(2),  483–518 (Jun 2023). \doi{10.1007/s40593-023-00340-7}

\bibitem{AYANWALE2022}
Adekunle, M., at~al.: Teachers’ readiness and intention to teach artificial intelligence in schools. Computers and Education: Artificial Intelligence  \textbf{3},  100099 (2022). \doi{10.1016/j.caeai.2022.100099}

\bibitem{bloom2001}
Anderson, L.W., et~al.: A Taxonomy for Learning, Teaching, and Assessing: A Revision of Bloom's Taxonomy of Educational Objectives (01 2001)

\bibitem{Christiansen2009}
Beckner, C., et~al.: Language is a complex adaptive system: Position paper. Language Learning  \textbf{59}(s1),  1--26 (2009). \doi{10.1111/j.1467-9922.2009.00533.x}

\bibitem{bergmann2012flip}
Bergmann, J., Sams, A.: Flip your classroom: Reach every student in every class every day. International society for technology in education (2012)

\bibitem{Bloom1968}
Bloom, B.S.: Learning for mastery. Evaluation Comment  \textbf{1}(2),  1--12 (March 1968), uCLA - CSEIP

\bibitem{Blossfeld2019}
Blossfeld, H.P., von Maurice, J.: Education as a Lifelong Process, pp. 17--33. Springer Fachmedien Wiesbaden, Wiesbaden (2019). \doi{10.1007/978-3-658-23162-0_2}

\bibitem{brown_adler_2008}
Brown, J.S., Adler, R.P.: Minds on fire: Open education, the long tail, and learning 2.0. EDUCAUSE Review  \textbf{43}(1),  16--32 (January/February 2008)

\bibitem{Christiansen2008}
Christiansen, M.H., Chater, N.: Language as shaped by the brain. Behavioral and Brain Sciences  \textbf{31}(5),  489–509 (2008). \doi{10.1017/S0140525X08004998}

\bibitem{Pevzner2015}
Compeau, P., Pevzner, P.A.: Life after moocs. Communications of the ACM  \textbf{58}(10),  41–44 (sep 2015). \doi{10.1145/2686871}

\bibitem{edge2024}
Edge, D., et. al.: From local to global: A graph {RAG} approach to query-focused summarization (2024), \url{https://arxiv.org/abs/2404.16130}

\bibitem{Engelbart1962}
Engelbart, D.C.: Augmenting human intellect: A conceptual framework. Tech. Rep. AFOSR-3223, SRI International, Menlo Park, CA (October 1962)

\bibitem{Fausey2010}
Fausey, C.M., Long, B.L., Inamori, A., Boroditsky, L.: Constructing agency: The role of language. Frontiers in Psychology  \textbf{1} (2010). \doi{10.3389/fpsyg.2010.00162}

\bibitem{Gligorea2023}
Gligorea, I., Cioca, M., Oancea, R., Gorski, A.T., Gorski, H., Tudorache, P.: Adaptive learning using artificial intelligence in e-learning: A literature review. Education Sciences  \textbf{13}(12) (2023). \doi{10.3390/educsci13121216}

\bibitem{Hou2017}
Hou, S.I., Pereira, V.: Measuring infusion of service-learning on student program development and implementation competencies. Journal of Experiential Education  \textbf{40}(2),  170--186 (2017). \doi{10.1177/1053825917699518}

\bibitem{hu2024}
Hu, Y., et. al.: Grag: Graph retrieval-augmented generation (2024), \url{https://arxiv.org/abs/2405.16506}

\bibitem{Jordan2014}
Jordan, K.: Initial trends in enrolment and completion of massive open online courses. The International Review of Research in Open and Distributed Learning  \textbf{15}(1),  133--160 (2014). \doi{10.19173/irrodl.v15i1.1651}

\bibitem{kahneman}
Kahneman, D.: Thinking, fast and slow. MacMillan (2011)

\bibitem{intro_to_llms_2024}
Karpathy, A.: Intro to large language models (2024), \url{https://www.youtube.com/watch?v=zjkBMFhNj_g}, accessed: 2024-09-23

\bibitem{KASNECI2023102274}
Kasneci, E., at~al.: {ChatGPT} for good? {On} opportunities and challenges of large language models for education. Learning and Individual Differences  \textbf{103},  102274 (2023). \doi{https://doi.org/10.1016/j.lindif.2023.102274}

\bibitem{Kolb2015}
Kolb, D.: Experiential Learning: Experience as the Source of Learning and Development (01 2015)

\bibitem{krinkin2024}
Krinkin, K.: Back to evolutionary intelligence: Reading {L}andgrebe and {S}mith. Cosmos+Taxis  \textbf{12}(5+6),  76--79 (2024)

\bibitem{krinkin2023}
Krinkin, K., Shichkina, Y.: Cognitive architecture for co-evolutionary hybrid intelligence. In: Artificial General Intelligence: 15th International Conference, AGI 2022, Seattle, WA, USA, August 19–22, 2022, Proceedings. p. 293–303. Springer-Verlag, Berlin, Heidelberg (2023). \doi{10.1007/978-3-031-19907-3_28}

\bibitem{krinkin2021}
Krinkin, K., Shichkina, Y., Ignatyev, A.: Co-evolutionary hybrid intelligence. In: 2021 5th Scientific School Dynamics of Complex Networks and their Applications (DCNA). pp. 112--115 (2021). \doi{10.1109/DCNA53427.2021.9587002}

\bibitem{krinkin2017}
Krusche, S.e.a.: Chaordic learning: A case study. In: 2017 IEEE/ACM 39th International Conference on Software Engineering: Software Engineering Education and Training Track (ICSE-SEET). pp. 87--96 (2017). \doi{10.1109/ICSE-SEET.2017.21}

\bibitem{lumsdaine1960}
Lumsdaine, A., Glaser, R. (eds.): Teaching Machines and Programmed Learning, A Source Book. National Education Association, Washington, D.C. (1960), eRIC Number: ED019872

\bibitem{minaeeLargeLanguageModels2024}
Minaee, S., Mikolov, T., Nikzad, N., Chenaghlu, M., Socher, R., Amatriain, X., Gao, J.: Large {{Language Models}}: {{A Survey}}

\bibitem{Morris2020}
Morris, T.H.: Experiential learning – a systematic review and revision of kolb’s model. Interactive Learning Environments  \textbf{28}(8),  1064--1077 (2020). \doi{10.1080/10494820.2019.1570279}

\bibitem{bjet.13121}
Nørgård, R.T.: Theorising hybrid lifelong learning. British Journal of Educational Technology  \textbf{52}(4),  1709--1723 (2021). \doi{10.1111/bjet.13121}

\bibitem{pinker2010}
Pinker, S.: The cognitive niche: {Coevolution} of intelligence, sociality, and language. Proceedings of the National Academy of Sciences  \textbf{107},  8993--8999 (May 2010), publisher: Proceedings of the National Academy of Sciences

\bibitem{bjet13123}
Poquet, O., de~Laat, M.: Developing capabilities: Lifelong learning in the age of ai. British Journal of Educational Technology  \textbf{52}(4),  1695--1708 (2021). \doi{10.1111/bjet.13123}

\bibitem{pressey1963}
Pressey, S.L.: Teaching machine (and learning theory) crisis.  \textbf{47}(1), ~1--6. \doi{10.1037/h0047740}

\bibitem{krinkin2022}
Schmiedmayer, P.e.a.: Global software engineering in a global classroom. In: 2022 IEEE/ACM 44th International Conference on Software Engineering: Software Engineering Education and Training (ICSE-SEET). pp. 113--121 (2022). \doi{10.1145/3510456.3514163}

\bibitem{Schoenemann}
Schoenemann, P.T.: Evolution of brain and language. Language Learning  \textbf{59}(s1),  162--186 (2009). \doi{10.1111/j.1467-9922.2009.00539.x}

\end{thebibliography}

\end{document}